\documentclass[12pt,preprint]{aastex}
\usepackage{psfig}
\newcommand{\sso}{$^{-1}$}

\begin{document}

\title{The BIMA Survey of Nearby Galaxies (BIMA SONG): \\
	I. The Radial Distribution of CO Emission in Spiral Galaxies}
\author{Michael W. Regan\altaffilmark{1,2,3} Michele D. 
	Thornley\altaffilmark{4,5}, Tamara T. Helfer\altaffilmark{6,7,8},
	Kartik Sheth\altaffilmark{3,9}, Tony Wong\altaffilmark{8},
	Stuart N. Vogel\altaffilmark{9}, 
	Leo Blitz\altaffilmark{8},
	and Douglas C.-J. Bock\altaffilmark{8}}
\altaffiltext{1}{Space Telescope Science Institute, 3700 San Martin Drive,
	Baltimore, MD 21218, mregan@stsci.edu}
\altaffiltext{2}{Carnegie Institution of Washington, 
	Department of Terrestrial Magnetism}
\altaffiltext{3}{Visting Astronomer, Kitt Peak National Observatory,
National Optical Astronomy Observatories, which are operated by the
Association of Universities for Research in Astronomy, Inc. (AURA)
under cooperative agreement with the National Science Foundation}
\altaffiltext{4}{National Radio Astronomy Observatory, Charlottesville, VA 22903}
\altaffiltext{5}{Physics Department, Bucknell University, Lewisburg, PA 17837}
\altaffiltext{6}{National Radio Astronomy Observatory, Tucson, AZ 85721}
\altaffiltext{7}{Steward Observatory, University of Arizona, Tucson, AZ 85721}
\altaffiltext{8}{Radio Astronomy Laboratory, University of California, 
	Berkeley, CA 94720}
\altaffiltext{9}{Department of Astronomy, University of Maryland, 
	College Park, MD 20742}

\begin{abstract}

We present the first results of
the BIMA Survey of Nearby Galaxies (BIMA SONG), an
imaging survey of the CO J=(1$-$0) emission in 44 nearby spiral
galaxies at a typical resolution of 6\arcsec.  
BIMA SONG differs from previous high-resolution CO surveys in that: 
(1) CO brightness was not
an explicit selection criterion; (2) a larger area (200\arcsec\ diameter
for most galaxies) of each galaxy was imaged; and (3) fully-sampled
single-dish CO data (55\arcsec\ resolution) were obtained for over
half of the sample galaxies, so
all of the CO flux is imaged in these galaxies.
Here we present CO maps for a subsample of 15 BIMA SONG galaxies 
for which we have also obtained near-infrared or optical broad-band
data.  
The CO maps display a remarkable variety of molecular gas
morphologies, and, as expected, the CO surface brightness distributions
show considerably more sub-structure than the stellar light distributions,
even when averaged over kiloparsec scales.    
The radial distribution of stellar light in galactic disks is generally
characterized as an exponential.
It is, therefore, of interest
to investigate whether the molecular gas, which is the star-forming
medium, has a similar distribution.
Though our low-resolution single-dish radial profiles of CO emission 
can be
described by simple exponentials, this is not true
for the 
emission at our full 6\arcsec\ resolution.
The scale lengths of the CO disks are correlated
with the scale lengths of the stellar disks with a mean
ratio of the scale lengths of about one. 
There is, however, considerable
intrinsic scatter in the correlation.
We also find that:
(1) there is also a weak correlation between the ratio of K-band to CO
luminosity and Hubble type;
(2) in half of the galaxies presented here, CO emission
does not peak at the location of the stellar nucleus;
(3) averaged over
the inner kiloparsec, the CO emission in one-half of the galaxies
exhibits an excess over that expected from
an exponential disk which is similar to
the excess in stellar light caused by the bulge stars;
and (4) this excess CO emission may be due to an increase in the total
molecular gas content in the bulge region, or alternatively, to an increase 
in the CO
emissivity caused by the increased pressure of the bulge region.

\end{abstract}

\section{Introduction}

The evolution of a spiral galaxy is intricately linked to its
molecular gas through a variety of processes.  
All available evidence suggests that
molecular gas fuels star formation both in the 
Milky Way \citep[and references therein]{Blitz93} and in other galaxies
 as can be seen through its
importance to studies of triggered star formation in spiral arms \citep{Vogel88},
nuclear starbursts \citep{YD91,Alto95,Planesas97}, and
the increased star formation seen at bar ends \citep{RTVV96,Sheth00}.
In addition, molecular gas can affect galaxy evolution through
inflow to the nuclear region (e.g. through bar or interaction driven inflow).
This inflow may contribute to non-stellar activity in the nucleus,
change the central mass concentration, cause a
bar to be destroyed \citep{Norman96}, or trigger a late Hubble type galaxy to evolve
into an earlier type \citep[e.g.,][]{FriedliMartinet93}.

The bulk of our knowledge about molecular gas in galaxies comes from
studies of carbon monoxide (CO) emission, primarily in its lowest
rotational transition (CO J=1$-$0, $\lambda$=2.6mm). 
CO is used as a 
tracer of H$_2$, which lacks a permanent dipole moment 
and is therefore undetectable at the
temperatures of typical molecular clouds (see review of CO
observations in galaxies by Young \& Scoville 1991).
Observations of CO have been made in
several hundred external galaxies \citep[for data prior to 1990 see
catalogs in][]{Verter85,Verter90}, with the largest uniform study
being the FCRAO Extragalactic CO Survey \citep{Young95}, which
presented spectra from 1412 positions in 300 galaxies.  Such observations have
served as the basis for our statistical understanding of the molecular
content of galaxies as a function of parameters like Hubble type and
luminosity. 

However, the distribution and physical conditions of
molecular gas on small scales within galaxies are still poorly understood.  
The overwhelming majority of extragalactic
observations have been carried out using single-dish telescopes, with
linear resolutions which are typically many kiloparsecs.  
Observations at this resolution resolve structures which are about one hundred
times larger in diameter than a giant molecular cloud (GMC), the basic 
unit into which molecular gas is organized in the Milky Way.  
Outside of the Local Group to 
achieve the sub-kiloparsec resolution needed to study the distribution
of molecular gas in galaxies, it is necessary to use an interferometer
for all but the nearest galaxies.
So far, the largest interferometric
study of external galaxies has been the recent survey presented by
\citet{Sak99a}, which mapped the central CO distribution in 20 nearby
spiral galaxies with the Nobeyama Radio Observatory (NRO) and the
Owens Valley Radio Observatory (OVRO) millimeter interferometers. 
The \citet{Sak99a} survey achieved 4\arcsec\ resolution over the central
region, within 30\arcsec\ of the center, 
and focused on how the distribution of molecular gas in the
central kiloparsec differs in galaxies with and
without strong nuclear activity \citep{Sak99b}.

The limited field of view and possible ``missing flux''
of the \citet{Sak99a} and other interferometer-only 
surveys \citep{Jogee99,Regan99} 
prevent us from making
a quantitative comparison of the molecular and stellar distributions
in the galaxies studied.
Meanwhile, the resolution of the single dish surveys
is not high enough to investigate the connection between features
in the stellar distribution (bars, spiral arms, and bulges) and the
molecular distribution.
To accurately investigate the sub-kiloparsec-scale
molecular gas distributions in galaxies,
we have undertaken the BIMA Survey of Nearby Galaxies (BIMA SONG), 
a systematic imaging survey of CO emission in the centers and inner
disks of an objectively selected sample of 44 nearby spiral galaxies.
The database we are producing from the survey includes spatial-velocity
data cubes showing the distribution and kinematics of CO emission at
resolutions of a few hundred parsecs ($\sim$6\arcsec)  
and $\sim$10 km~s$^{-1}$ over a field of view of
typically 10 kiloparsecs ($\sim$190\arcsec).  The BIMA SONG maps for
over half of the 44 sample galaxies incorporate single dish data taken
with the NRAO 12m telescope; these maps, therefore, do not suffer from
the ``missing flux'' problem to which interferometric images are often
susceptible.  For the remaining galaxies, we collected sensitive
spectra from the NRAO 12m telescope so that we could assess how much, if
any, of the single dish flux is missing in the BIMA-only maps.
BIMA SONG differs from previous
high-resolution CO surveys in the following ways: (1) CO brightness was
not an explicit selection criterion, as we observed all nearby, optically-bright
spirals with suitable declinations and inclinations (except M33); 
(2) we observed a much larger area of each galaxy, covering a
significant fraction of the optical disk; 
(3) we observed all sample galaxies using uniform criteria, and reduced
and analyzed all the data using well-defined, uniform procedures; and 
(4) we incorporated fully-sampled single-dish CO data into
the BIMA SONG maps for over half of the galaxies, 
so that all CO flux is included for these galaxies.

In this paper, we address one fundamental aspect of the molecular gas
distribution in galaxies: what is the radial surface brightness
distribution of molecular gas, and how does it compare to that of the
stars? 
Because stars form from molecular gas, the molecular
disk in a galaxy and the stellar disk, where the bulk of the new
stars reside, should
have related radial distributions.
In fact, \citet{YS82} used single dish observations to show that,
in the two galaxies they studied, the single dish radial profile and
the optical B-band profile had very similar scale lengths.
On the other hand, \citet{Sak99a} derived a 1/$e$ length of only
500 pc for their sample; this is a small fraction of the stellar light
scale length.
\citet{Sak99a} compared their data with results from single-dish
studies of the same galaxies and suggested that they were detecting a central CO
concentration distinct from the large-scale gas disk. 

In order to compare the radial distributions of molecular gas and star formation
over a range of radii and size scales, 
we have chosen a subsample of 15 BIMA SONG galaxies
for which we currently have near infrared or optical images
and for which we have incorporated single-dish data into the BIMA SONG
maps.  We find that the CO distributions 
for most of the galaxies show
significant deviations from the stellar disks on both large
($\sim$60\arcsec\ or more) and small
($\sim$6\arcsec) scales.  
We also find that in over half of the
galaxies, the CO surface brightness increases in the central kiloparsec in a
similar fashion to the increase in the stellar surface brightness
caused by the bulge stars.  This
increase may either represent a true increase in the molecular mass
or it may instead represent an increase in the emissivity of the CO.

\section{BIMA SONG Sample Selection}\label{Ssongselect}

We selected galaxies for the complete BIMA SONG sample
using criteria which are not explicitly
based on the CO luminosity of the galaxies.  The galaxies were
selected to have Hubble
types between Sa and Sd with a recessional velocity $v_\odot < 2000$
km~s\sso; if $H_0$=75 km s$^{-1}$ Mpc$^{-1}$, this corresponds to
galaxies with Hubble distances of $d \la 27$ Mpc.  The galaxy
inclinations are all $<$70\arcdeg (based on axial ratios)
so that we may study properties
which vary azimuthally.  A minimum declination of $-$20\arcdeg\ ensures
that the sources can be observed well from the Hat Creek Radio
Observatory in northern California.  Finally, to limit the size of the
sample, we chose galaxies with apparent blue magnitudes brighter than
11.0. 
Using the NASA/IPAC Extragalactic
Database (NED)\footnote{NED is operated by the Jet Propulsion
Laboratory, Caltech, under contract with the National Aeronautics and
Space Administration.} we identified 45 galaxies meeting these
criteria.  All except M33 were observed, yielding a sample size of 44
galaxies for BIMA SONG.  At
the average distance of the sample ($\sim$11 Mpc), our nominal
synthesized beam of 6\arcsec\ provides a linear resolution of 330 pc.
Table \ref{sampletab} lists the galaxies selected for the BIMA SONG
survey.

For this paper, we selected galaxies for which we
have an optical or near-infrared image as
well as a single dish
CO map from the NRAO 12m telescope. 
The subsample (Table \ref{galparm}) tends to be biased towards the brighter CO
galaxies in the full BIMA SONG sample, but its distribution in
Hubble type (Sab-Scd) and bar type (7 or 47\% are classified as SAB or
SB, vs. 24 or 54\% in the full sample) is representative of 
the full sample.  None of the galaxies in this first subsample is
significantly distorted or asymmetric, as might be expected in
strongly interacting systems.  The subsample does include galaxies
which are weakly interacting or have smaller companions, such as NGC 3627, NGC
5055, NGC 5194 (M51), and NGC 7331.

\section{Observations}\label{Sobserve}

\subsection{BIMA Observations}

We carried out
BIMA SONG observations from November 1997 through
June 1999 using the 10-element Berkeley-Illinois-Maryland Association
(BIMA) millimeter interferometer \citep{Welch96} at Hat Creek, CA.
We made the observations using the C and D array configurations,
which include baselines as short as 7.6m (with the 6.1m BIMA antennas as 
closely spaced as possible) and as long as 90m.  
We observed
two-thirds of the galaxies
(those with $R_{25}$ $>$ 200\arcsec) using a 7-field,
hexagonal mosaic with a spacing of 44\arcsec; 
this pattern yields a
half-power field of view of about 190\arcsec, or 10 kpc, at the average
distance of galaxies in the survey.  
We observed
the remaining galaxies 
with a single pointing, which yielded a field of view of
100\arcsec\ FWHM.  
We observed
some galaxies (NGC~0628, NGC~1068, NGC~2903,
NGC~3627, NGC~4736 \citep{WB00}, NGC~5033 \citep{W00}, \& NGC~5194) with slightly
different pointing spacings or additional fields.  
For the
multiple-pointing observations, we observed each field for one minute
before we switched to the next pointing.  
Thus, we returned to each
pointing after $\sim$8 minutes, which is well within the time needed
to ensure excellent sampling of even the longest baselines in the $uv$
plane.  (For a 100\arcsec\ source, the time to cross an individual
$uv$ cell at 6\arcsec\ resolution is about 48 minutes; see Welch
1993.)  
We configured
the correlator to have a resolution of 1.56 MHz
(4 km s\sso) over a total bandwidth of 368 MHz (960 km s\sso).

\subsection{ NRAO 12m Observations }\label{12mobs}

We collected single-dish data over several observing seasons from 1998
April through 2000 June using the NRAO\footnote{The National Radio
Astronomy Observatory is operated by the Associated Universities,
Inc., under cooperative agreement with the National Science
Foundation.} 12m telescope on Kitt Peak, AZ.  
We observed orthogonal
polarizations using two 256 channel filterbanks at a spectral
resolution of 2 MHz (5 km~s\sso), and using a 600 MHz configuration of
the digital millimeter autocorrelator with 0.8 MHz (2 km s\sso)
resolution as a redundant backend on each polarization.  
We monitored the pointing
every 1-2 hours with observations of planets and strong
quasars.  
We also measured the focus at the beginning of each session and
after periods during which the dish was heating or cooling.

In order to minimize relative calibration errors and pointing errors
across the map, we observed in On-the-Fly (OTF) mode \citep{Em96},
where the telescope takes data continuously as it slews across the
source.  
In this mode, the actual telescope encoder positions are read
out every 0.01 seconds and folded into the spectra, which are read out
every 0.1 seconds.  
Each 6\arcmin$\times$6\arcmin\ OTF map takes $<$20
minutes to complete, and each source needed 10 to 30 OTF maps 
to achieve the desired sensitivity.  (Following
\cite{Cornwell93}, we tried to spend enough time on the single-dish
measurements to match the signal-to-noise ratio of the interferometric
observations. In practice, the BIMA maps tended to have better noise
levels by a factor of about two)
Given reasonably stable observing conditions, the
relative flux calibration across an individual map will be very
good, so that even if the absolute flux scale drifts from map to map,
the combined final 12m map will have very good pixel-to-pixel
calibration. 
A careful analysis of well-pointed data (\S\ref{pxcor})
over many observing seasons shows that the absolute flux calibration
at the 12m is accurate to better than 10\%.  
We 
weighted and gridded the OTF data
to an 18\arcsec\ cell in AIPS. 

\subsection{Optical/Infrared Observations \& Calibration}

The optical or near-infrared images used in this paper were taken either
from the literature or from a set of observations which were
taken to form a complementary database of broad-band images for SONG
galaxies.  Table \ref{Toptir} summarizes the data used
for each of the galaxies presented in this paper.

We observed
NGC 3351 and NGC 4321 using the Dupont 2.5m at Las
Campanas in May of 1998 in non-photometric conditions.
We used the wide field reimaging camera with the TEK5 chip yielding a
plate scale of 0\farcs7 per pixel.
These were calibrated
using published aperture photometry values.  
We observed NGC 4258 and NGC 4736 using the 0.9m at Kitt Peak
in April of 1999 in photometric conditions.
We used the T2KA chip in direct imaging mode, which yielded a plate
scale of 0\farcs38 per pixel.
Finally, we observed NGC 4826 using the 1.5m at Palomar Observatory
in April of 2000 in photometric conditions.
The plate scale of the images after using on-chip 2x2 binning was 
0\farcs78 per pixel.
For all of the observing runs, 
the data were flat fielded, cosmic ray rejected, and sky subtracted
using standard IRAF routines.

\section{CO Data Reduction and Calibration}\label{Sreducs}

\subsection{Reduction of BIMA data}

We conducted
the BIMA CO data reduction using standard tasks available in the MIRIAD package
\citep{Sault95}.  
We removed the instrumental and atmospheric phase
variations from the source visibilities by referencing the phase to
observations of a nearby quasar every 30 minutes. 
We measured and removed the 
antenna-based amplitude and phase variations as a
function of frequency, using the BIMA online
passband observations at the time of each track.
After inspecting our observations of our primary flux calibrators 
for any residual, baseline-based
passband errors, we concluded that no further passband corrections were
necessary.
  
We detected continuum emission in NGC~1068 and NGC~3031. 
For these sources, we subtracted the continuum in the visibility
plane by fitting a constant to the real and imaginary
parts of those channels in the passband that were free of line
emission.  The continuum maps derived in this way were consistent with
maps made from the line-free lower sideband of the local oscillator at
112 GHz.

Additional steps in the data reduction process for BIMA SONG data,
customized to ensure uniformity across the BIMA SONG sample, are 
described in the following subsections. A detailed description of specific imaging and
data combination techniques will be presented in \citet{Helfer01}.

\subsubsection{Atmospheric Decorrelation Correction}

Differences in the column of water vapor between a pair of antennas are
a source of phase noise that increases with increasing baseline
length. 
This phase noise results in a decorrelation or reduction in the
measured visibility amplitude, which distorts the high-resolution
map structure.  
We made 
estimates of the decorrelation using
measurements taken by the  real-time phase monitor \citep{Lay99} at Hat Creek.
The phase monitor reports the rms phase variation
over a fixed baseline of 100m every 10 minutes
based on observations of a
commercial direct broadcast satellite. 
We scaled the phase monitor data to our
observing frequency, assumed that the phase rms scales as (baseline
length)$^{5/6}$, assuming Kolmogorov turbulence \citep{Kolm41,Akeson99}, 
and corrected for the airmass of
each visibility.   
In general, data for which the decorrelation was large
were rejected and not included in the final datasets.  
For the included datasets,
we decreased the weight of
the data according to the estimated loss in sensitivity resulting from
decorrelation and scaled the visibility amplitude up according to
the estimated decorrelation.
This technique results in a
dirty map which is essentially unchanged, but a dirty beam that more accurately
represents the instantaneous atmospheric-limited response of the 
interferometer.
It also enables a more accurate deconvolution,
as the final beam size more accurately represents the degraded
resolution, and produces a better estimate of the true noise level.
The typical increase in
beam size was only $\la$ 5\%, reflecting the high quality of 
most of the data.

\subsubsection{Amplitude Calibration\label{coampcal}}

We set the amplitude scale for the BIMA SONG maps using observations
of the nearby quasar that was used for phase calibration.  
Since the flux of quasars are generally variable, we determined their fluxes by comparing
them to the primary flux calibrators Mars or Uranus.  
In addition, we used 3C273 as a secondary calibrator.
We observed one of
these flux calibrators in almost every track.  
While the brightness
temperature for the planets is generally well-determined, the flux of
3C273 is variable.  
To determine the flux of 3C273, we extracted
observations of 3C273 from the BIMA archives and calculated the flux
of 3C273 every time it was observed with a planet in any track at any
frequency in the 3mm band during the period of our observations.  We
estimate the uncertainty in our determination of the flux of 3C273 to
be less than 10\%.  

With these uncertainties in calibrator fluxes, measured and estimated errors
in the gains at 115 GHz, and estimates of atmospheric decorrelation on
time scales shorter than 10 minutes, we assign a one sigma
uncertainty to our amplitude calibration of 15\%.

\subsubsection{Self-Calibration}

Although the use of a nearby quasar to calibrate phase variations
removes most of the phase errors in the observations, there can still
be errors introduced by the atmospheric difference between the phase
calibrator and the source and by incorrect baseline determinations.
To correct for these errors, we performed an iterative phase-only
multi-channel self-calibration on our source using the initial 
phase-calibrated map as the model for the first iteration.
The model output from the deconvolution of this map was used as the
input for the self-calibration, and the self-calibrated visibilities
were then used to make a new map.  This process was repeated for three
iterations.  In general, the phase gains determined by this process
were small ($<$20\arcdeg) and the maps were not greatly changed.  We
checked that the source position was not affected by self-calibration
by comparing the position of the source in 
the self-calibrated map to its position in the {\it a priori}
phase-calibrated map.
  
\subsubsection{Production of BIMA CO Maps}

In producing initial maps of the CO emission, we weighted the visibility data by
the noise variance to account for differences in system temperatures
and gains of the individual antennas.  We further applied robust
weighting \citep{Briggs95,BSS99,CBB99} to the visibilities.  We
included shadowed data down to projected separations of 5m.
For each galaxy, we smoothed the amplitude- and phase-calibrated BIMA
data to 10 km~s$^{-1}$ resolution (a few narrow-line sources were smoothed to 5
km~s$^{-1}$) and generated data cubes in right ascension,
declination, and LSR velocity.  We deconvolved the BIMA-only data
cubes using a Steer-Dewdney-Ito CLEAN algorithm \citep{SDI84}.  These
BIMA-only maps were used to register pointing of the single-dish OTF
maps (\S\ref{pxcor}) 
prior to combining the BIMA and single-dish data cubes (\S\ref{Scombine}).

\subsection{Combination of BIMA and Single Dish Data}

We have investigated several methods for combining the interferometer and
single dish data in order to produce maps with no ``missing'' flux; 
the results of this analysis will be presented by
\cite{Helfer01}.  
In this initial paper, we present maps made using a
linear combination method described in \cite{Stan99}. 
The following subsections describe the production of these maps.

\subsubsection{ Pointing Cross-correlation }\label{pxcor}

The implementation of OTF mapping at the NRAO 12m
(\S\ref{12mobs}) ensures that the internal pointing consistency of an
individual map (taken over $\sim$20 min) is excellent, even if the
overall registration of the map is uncertain.  We were further able to
cross-correlate either the individual OTF maps or averages of several
maps with each other to track relative pointing shifts, allowing OTF maps
taken over many hours, days, or even in different observing seasons to
have good internal pointing accuracy.  The absolute registration of
the 12m maps was accomplished by cross-correlating the well-pointed
average 12m map with the BIMA map, where the BIMA map was smoothed to
match the 55\arcsec\ resolution of the 12m at 115 GHz.  (In performing
this cross-correlation, we must assume that the flux resolved out in
the interferometer maps does not affect the centroid of the emission.
A theoretically more robust method, comparing flux density just in the
region of $uv$ overlap, is limited by poor signal-to-noise ratios.)  
We used the pointing
cross-correlation to correct the registration of the 12m maps.
In general, these corrections were 5-10\arcsec\ but some were
as large as $\sim$20\arcsec. 

\subsubsection{Interferometer and Single Dish Combination\label{Scombine}}

Using the \cite{Stan99}
method, we created a new ``dirty'' map by a linear combination of the
interferometer map (prior to CLEANing) and the single dish map, 
with the latter tapered
by the primary beam response of the BIMA map (for a mosaic this 
is the combined response of the individual pointings).
In performing this combination, the single dish map was downweighted
according to the ratio of the area of the BIMA beam to the area
of the 12m beam (typically about 0.01), and a new dirty beam was created
by linearly combining the synthesized BIMA beam and the 12m beam (assumed
to be a truncated Gaussian) using the same relative weighting.

We then deconvolved
the   new combined ``dirty'' map
in the usual way using a Steer-Dewdney-Ito
CLEAN algorithm.
Hereafter, the cleaned maps resulting from this technique are called
``BIMA+12m'' or combined maps.

\subsubsection{Production of Integrated Intensity Maps}

To create images that best show the total integrated emission from
each galaxy, we formed moment maps using a smoothed version of the
data cube as a mask.  
We formed the mask
by smoothing the
data by a Gaussian with FWHM=20\arcsec\ and then accepting all
pixels in each channel map where the emission in the smoothed data cube
was above three times the
noise. (We calculated the noise in emission-free channels of the
smoothed cube).  
The moment map was then the sum of all emission in the
accepted pixels of the full-resolution cube, multiplied by the
velocity width of an individual channel.  
Although this
technique is sensitive to low-level emission distributed
similarly to the bright emission, it introduces a bias against
compact, faint emission which is distributed differently than the
brightest emission.
Figure \ref{moments} (left
hand column) shows the relationship between the CO and stellar
morphologies by overlaying the CO emission distributions on our near
infrared/optical images of the galaxies.  
The total emission (moment 0) images for the galaxies are shown
separately in Figure \ref{moments} (right hand column).  
For all quantitative
analyses in this paper, we used moment maps made without any masking
 to avoid the potential bias described above.
\section{Results}

\subsection{Molecular Gas Distributions}\label{molgasdist}

The 15 maps shown in Figure \ref{moments} (right hand column) reveal a
variety of molecular morphologies that are representative of the BIMA
SONG sample as a whole.  
The FWHM of the synthesized beams is
indicated in the lower left corner of the images in
Figure \ref{moments} and is listed in Table \ref{galparm}.
Structures are present in these maps at the smallest scales sampled by 
the observations, typically about 6\arcsec, as
well as on scales many times larger than the synthesized beams.  There
are grand-design spiral galaxies (NGC~0628, NGC~5194), where the CO is
tightly confined to the spiral arms.
These arms wind into the nuclear region, which lacks a single nuclear peak.  
By contrast, the molecular
gas appears to be more smoothly distributed in the flocculent galaxies NGC~4414
and NGC~5055; this is at least partially due to the higher inclination
of these galaxies, which produces a lower linear
resolution parallel to the minor axis.  
The barred galaxies show a diversity of structure: in
NGC~3351, the gas appears concentrated nearly entirely along the
``twin peaks'' area of the nuclear region \citep{Ken92}, whereas in
NGC~3627, molecular gas is observed in a central concentration, along
the leading edges of the bar, at the end of the bar, and extending
over nearly 4\arcmin\ on the sky in spiral arms that extend from the
ends of the bar.  Although the highest CO surface brightness often
occurs at or near the nucleus of the SONG galaxies, this is not
universally the case: NGC~3521, NGC~4414, and NGC~7331 have a
pronounced lack of emission at their centers. Furthermore, in NGC~628, 
NGC~1068, NGC~3351, NGC~4258, and NGC 5194 the strongest
emission is not associated with the nucleus of the galaxy.

\subsection{Radial Profiles of Molecular Gas}

To explore the radial distributions of molecular gas, 
we calculated CO radial profiles from
 the $\sim$6\arcsec\ resolution
BIMA+12m combined maps as well as the 55\arcsec\ resolution 12m OTF maps.
We determined the position angles and inclinations (Table \ref{galparm})
from either the stellar light or from the CO kinematics 
and then measured the
average brightness in concentric elliptical annuli for each source.
We fixed the centers for all profiles at the position of the peak in the
stellar maps.  We used unclipped moment maps
to avoid any biases against faint emission and measured the average in
each annulus.  Annuli were spaced at approximately
one-half of the resolution of the corresponding image.  Thus, the CO
BIMA+12m profiles are described by data points spaced 3\arcsec\ apart
and the 12m profiles are described by data points
spaced 27\arcsec\ apart.

We converted CO brightnesses to logarithmic units in order   
to compare with stellar profiles.  
However, the use of a standard magnitude scale leads to difficulties at radii
where the brightness approaches the noise level, since here the
brightnesses can be very small or even negative (e.g., due to residual
deconvolution errors).   
To put the data on a scale which is better behaved near the noise
level of the image, we adopted a
modified magnitude scale recently developed for optical studies.  
We used
inverse hyperbolic magnitudes \citep{LGS99} with the zero
magnitude set at 1000 Jy km s$^{-1}$ arcsecond$^{-2}$.  
In detail, we converted to magnitude $\mu$ using
\begin{equation}
\mu(x)=-2.5\log_{10}(e)\left[\sinh^{-1}\biggl(\frac{x}{2b}\biggr) +
	\ln{b}\right],
\end{equation}
where $x$ is the flux and $b$ is the ``softening'' parameter that
determines the threshold of linear behavior of the magnitudes.  
When $x$ is large relative to $b$, these magnitudes reduce to the standard
astronomical magnitude scale.  \citet{LGS99} showed that the optimum
value for the softening parameter $b$ is the noise level of the flux
measurement.  
As the noise level of the combined maps varied little from galaxy to galaxy,
the softening parameter was held constant at $b$=1.33 Jy km\sso s\sso beam\sso.  

The 55\arcsec\ resolution data from the 12m
telescope produce smooth, monotonically decreasing profiles that can be
described by a single exponential (with the possible exception of NGC
7331).  
The BIMA+12m 6\arcsec\ resolution maps, on the other hand,
yield complex, heterogeneous profiles, with a great deal of structure
on sub-kiloparsec size scales.

\subsection{Comparison of CO and Stellar Radial Profiles}

Stellar profiles were produced for the optical/infrared images of our
subsample galaxies in a manner similar to the CO profiles.
Due to the higher resolution available at these wavelengths, points on
the optical/infrared profiles are spaced 1\arcsec\ apart.
In addition, we used the median intensity
in each annulus to avoid surface brightness excursions caused by
foreground stars.

\begin{figure}
\epsscale{0.65}
\plotone{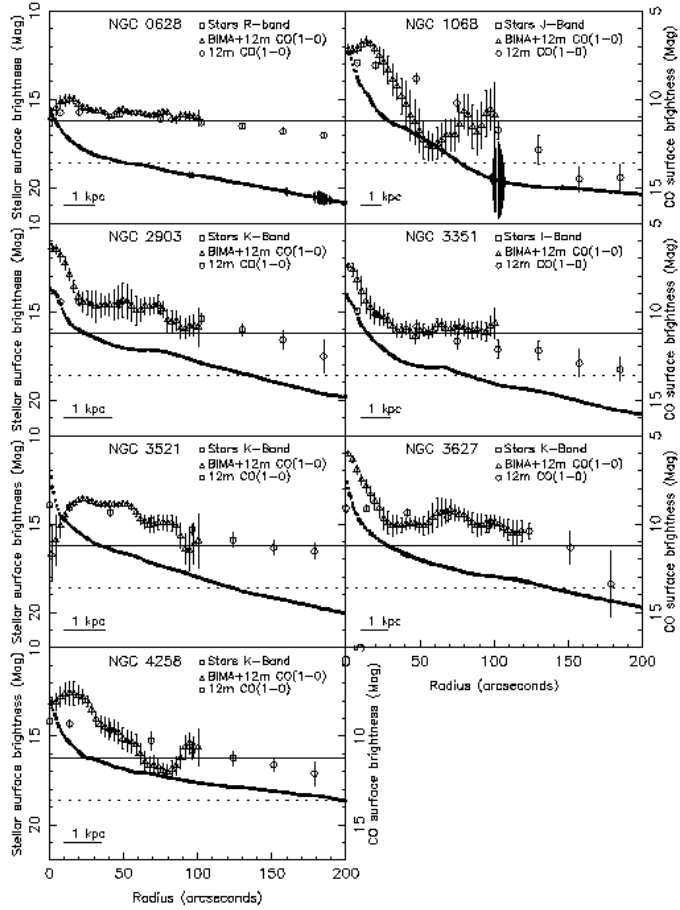}
\caption{
Radial profiles of the CO and stellar emission. The CO surface
brightness (triangles for BIMA+12m, circles for 12m only)
is in magnitudes of Jy km s\sso\ arcsec$^{-2}$
with the zero point in magnitudes defined to be at 1000 Jy km
s\sso\ arcsec$^{-2}$. 
The stellar surface brightness (squares) is in
K-band magnitudes arcsec$^{-2}.$  
All points show one sigma error
bars.  
The solid horizontal line is the noise-dependent flux level of the
BIMA+12m maps that is consistent with a zero flux.  Similarly, the dotted
horizontal line is consistent with a zero flux level in the single dish
only maps.  The horizontal bar in the lower left corner shows the
angular size of 1 kpc at the assumed distance to the galaxy.}
\label{coprofiles} 
\end{figure}

Figure \ref{coprofiles} compares the stellar profiles with the 
CO radial brightness distributions for each galaxy.  
The stellar profiles show the expected exponential disk plus a central 
de Vaucouleurs ($r^{1/4}$) bulge component. 
Most remarkably, in 8 of the 15
galaxies (see Table \ref{scaletab})
in Figure \ref{coprofiles}, there appears to be a separate,
bulge-like excess CO component to the BIMA+12m profiles that occurs
over a radial extent roughly similar to the bulge component of stellar
light.
Here we classify a galaxy as having an excess CO central component when the central
CO surface brightness is brighter than the extrapolation of the large scale
disk to zero radius.

It would be surprising if the CO emission and bulge star light were 
directly related
because the optical/IR bulge is the projection of a
3-dimensional distribution while the CO distribution throughout the
entire galaxy is almost surely a flattened disk.  Furthermore, the
BIMA+12m data are, in general, not well fit by an exponential radial
distribution even at radii where there is no contribution from the
nucleus.  
Apparently, the low resolution data often mask complex
radial profiles as well as large variations in the azimuthal
distributions.

\begin{figure}
\epsscale{0.75}
\plotone{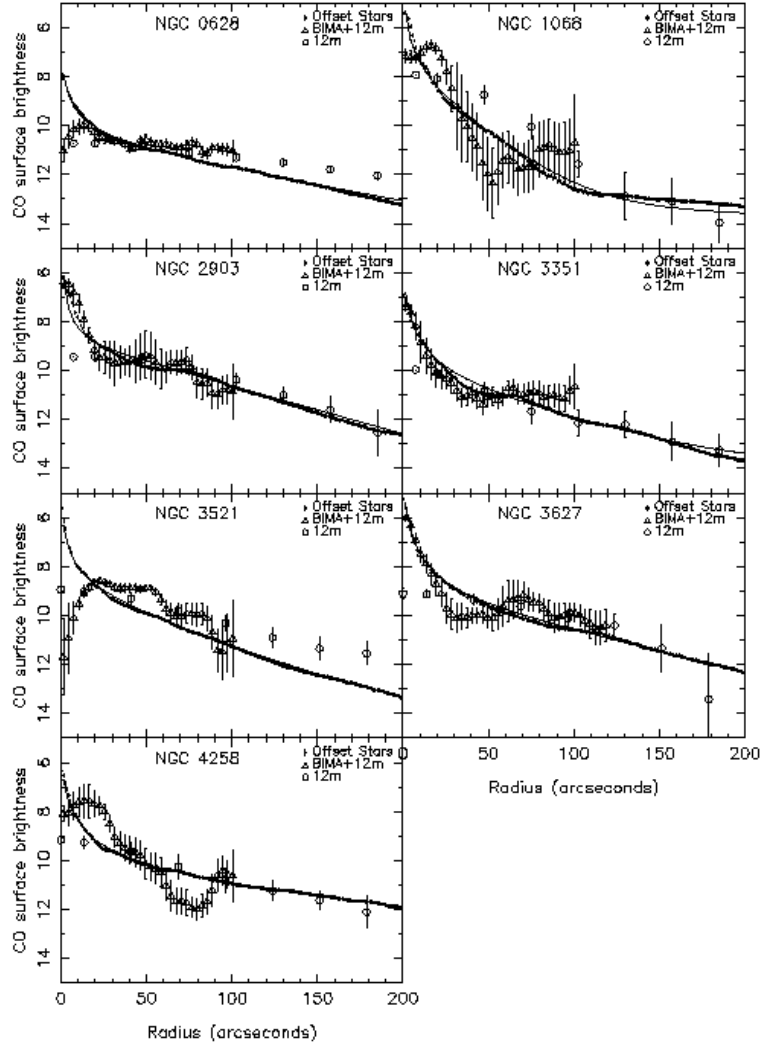}
\caption{
CO and offset stellar profiles. Similar to Figure \ref{coprofiles}
except that here the stellar profile has been offset to the average
value of the offset between the CO+12m and the stellar profile.
The
asterisks are the surface brightness values of the stellar profile after
offsetting. 
The solid line is the offset model profile.
The triangles are the surface photometry of the CO+12m
maps, and the circles are the surface photometry values from the
12m-only CO maps.}
\label{modprof}
\end{figure}

To compare the BIMA+12m CO surface brightness profiles and the stellar
surface brightness profiles directly, we determined the average
magnitude offset between the two profiles.
First, we convolved the stellar images
to the resolution of the CO images and 
sampled the stellar
images at the same intervals as those of the CO images to form a
matching stellar radial profile.  
We used this profile as input to a
joint bulge/disk decomposition routine \citep{RV94}.  Due to the low
resolution of the convolved stellar profile 
in the central region, a good fit is obtained without uniquely
determining the parameters of the individual bulge and disk components.  
From these model parameters, we then
created a model stellar profile to deemphasize the deviations in the
original stellar surface brightness profile caused by spiral arms,
bars, and foreground stars.  We determined the offset between the
stellar profiles and the BIMA+12m profile by averaging the
offsets at each annulus.  Figure \ref{modprof} shows a plot of the
offset modeled stellar surface brightness profiles, the offset
stellar profile, the CO BIMA+12m,
and the 12m profiles.  The figure shows that, although the stellar and
CO profiles display a similar surface brightness distribution in some
galaxies, in general, there are deviations of two magnitudes or
more over a large range in radius.  In some nuclei, this difference is
as much as four magnitudes.  

\subsection{Constancy of the CO to K-band Flux Ratios}\label{cokratio}

\begin{figure}
\plotone{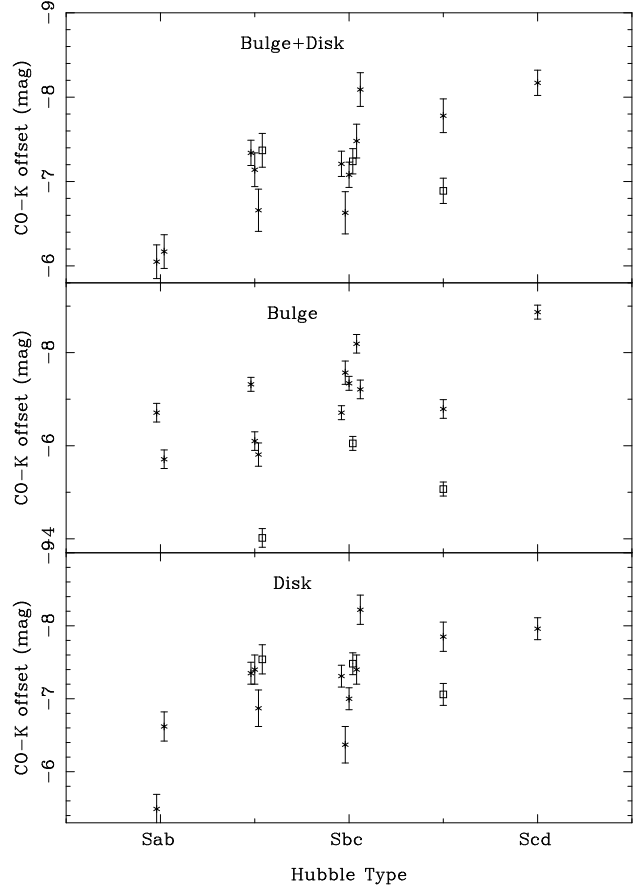}
\caption{
(Upper panel) CO-K offsets of the inner 200 arcsec.
The total CO flux is obtained from the
single dish map. The stellar observations that were not done in K-band
were corrected to K-band magnitudes using published aperture
photometry values.  The error bars are the quadrature sum of the
uncertainties in the K band magnitudes and those of the CO amplitude
calibration and are dominated by the uncertainties in the CO.  The
three galaxies plotted as boxes are NGC 4414, NGC 3521, and NGC 7331,
the three with strong central depressions in their surface brightness
profiles. 
(Middle panel) 
CO-K offsets of bulge region.  This is the offset of the light inside a
one kpc radius.  The bulge CO flux is obtained from the BIMA+12m
map.
(Bottom panel)
CO-K offset of disk region (r $>$ 1 kpc).  This is the offset of the light outside of
a one kpc radius.  The disk CO flux is obtained by subtracting the
bulge flux obtained from the BIMA+12m map from the single dish flux
inside of 200\arcsec. 
}
\label{alloffset}
\end{figure}

Is the qualitative agreement between the stellar
surface brightness profiles and the low resolution 
CO surface brightness profiles
evidence of a relationship between the CO flux and the stellar flux in
these galaxies?  For comparison, we converted all the stellar images
to a K-band magnitude scale using colors determined from published
aperture photometry values (see Table \ref{Toptir} for colors and
references).  
The uncertainties in this
calibration are the dominant source of uncertainties in the total
stellar magnitudes, but they are small compared to the uncertainties
of the CO flux scale (see \S\ref{coampcal}).  
We calculated the magnitudes using the total
flux inside a radius of 190\arcsec\ for the single dish CO and
stellar profiles.  Figure \ref{alloffset} (upper panel) 
shows the resultant CO-K
offset as a function of the Hubble type of the galaxies.  
Here a relatively brighter CO flux compared to the stellar flux
results in a more negative CO-K offset.
There is a trend of increasing CO luminosity relative to the
stellar luminosity for later Hubble types.
This increase in CO luminosity is the opposite of what was seen
by \cite{Casoli98}, who found that CO emission decreases relative
to dynamical mass with later Hubble types.

 The trend seen in the top of Figure \ref{alloffset} is a semi-quantitative
expression of gaseous properties of the Hubble sequence.  Late-type
galaxies contain more gas and dust than early-type galaxies \citep{sandage61}, 
and the more negative CO-K offsets in the later type galaxies
imply that, globally, these galaxies contain, on average, more molecular gas and,
by implication, more dust than the early-type galaxies.
Given that the HI surface densities saturate at an A$_v$ of 0.5 mag
(Wong 2000 and references therein), inclusion of HI, which is beyond
the scope of this paper, should add to the scatter in the relation and
perhaps affect the slope somewhat, but not the trend itself.

Figure \ref{modprof} shows that for 8 of the 15 galaxies, the CO
surface brightness increases in the bulge region in a manner similar
to the stellar surface brightness.
The presence or absence of a central excess is summarized in Table
\ref{scaletab}.
To investigate quantitatively 
whether the relationship between CO luminosity and
K-band flux is the same in  the bulge and disk regions,
we show the CO-K offsets of the bulge regions and disks
as a function of Hubble type
in Figure \ref{alloffset} (middle panel) and
(bottom panel), respectively.
We take one kiloparsec as
the fiducial radius of the bulge and sum all the flux inside this
radius, using the BIMA+12m for the CO.  The remaining flux out to a
radius of 190 arcseconds (from the single dish CO data) is
assigned to the disk component.
The flux values are shown in Table \ref{KCOmag}.

Linear regressions for all three plots in Figure \ref{alloffset} 
give similar slopes and intercepts. 
The slopes for the entire galaxy and the disk alone are --0.39$\pm$0.11 and
--0.37$\pm$0.14 respectively; the slope for the bulge alone is
--0.47$\pm$0.28 (with one sigma uncertainties).
The intercept for the entire galaxy is --5.7.
The fits show that there is a statistically
significant relationship between the CO-K offset and the
Hubble type.  
The K-band light is thus a 
predictor of the CO flux, not only in the disk but also in the bulge,
and the relationship between the two quantities is the same,
within the uncertainties, both in the bulge and the disk.  
This is quite surprising given the
large variation of emission distributions in the sample galaxies
shown and the difference in geometry between the stellar and CO
emission from the bulge region.

There is a significant difference between the amount of light 
contributed by disk stars and bulge stars in the two different regions.
As one would not expect the light from bulge stars to be related
to the CO emission, one would expect the CO-K offset of the bulge
region to be different from that of a region where the light is 
dominated by disk stars.
We can use our simple disk/bulge decomposition model to give us an
estimate of how much of the stellar 
light in the inner kiloparsec arises from
the bulge stars and how much comes from the disk.  
By averaging all
fifteen galaxies, we find that the disk-only stellar light is
2.2$\pm$0.14 magnitudes fainter than the total light within the inner
kiloparsec.  
Thus, if the CO surface brightness were related only to
stars in the disk, then the CO-K offsets in the bulge region would be
2.2 magnitudes larger than the mean of the values measured for the disks
in the bottom panel of Figure \ref{KCOmag}.
The mean of the points in the bulge, shown in the middle panel,
is, however, only 0.52$\pm$0.38 larger than what is measured for the 
disks of the galaxies, an increase of only marginal significance.
Thus, the mean CO-K offset is very nearly the same in the bulge as
in the disk, a rather surprising result given that the geometries
of the two stellar components are so different.

\subsection{ Scale Lengths of the CO and Stellar Disks }

\begin{figure}
\epsscale{0.75}
\plotone{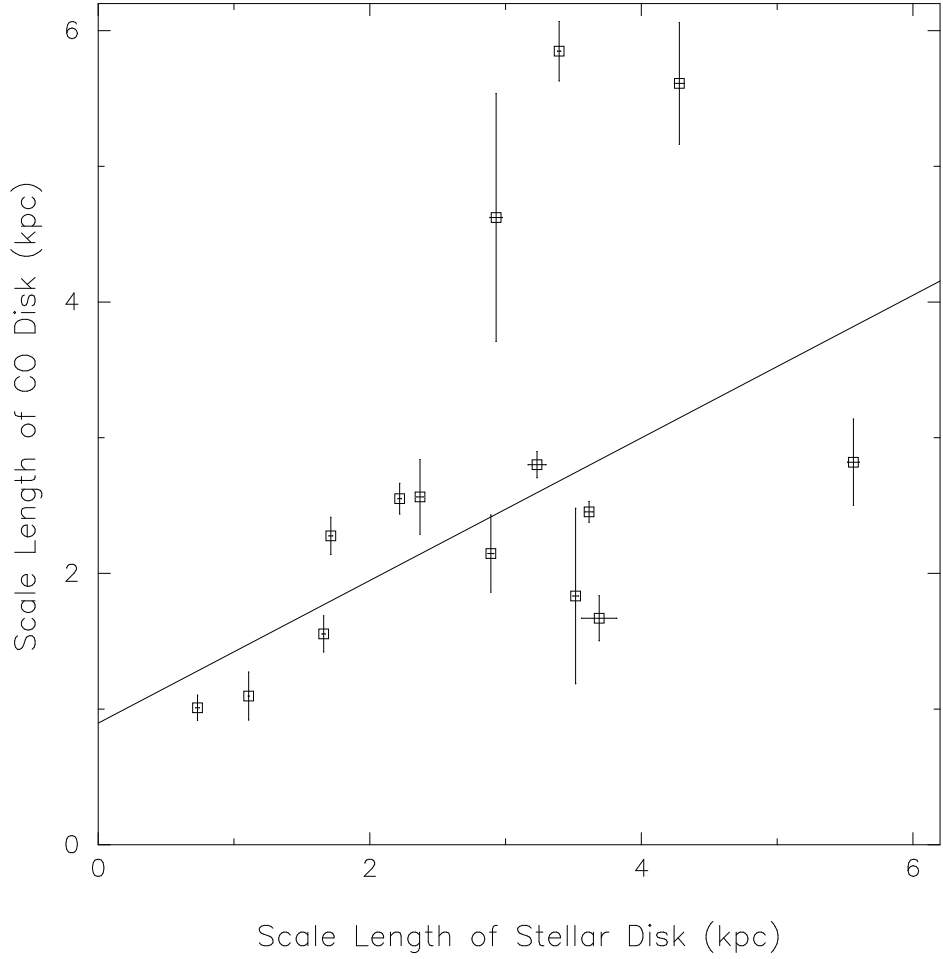}
\caption{
Scale lengths of CO and stellar disks. The error bars are the
formal one sigma uncertainties in the fitted scale lengths.
The CO scale lengths are derived from the fully sampled single dish maps.
The solid line represents a linear
least-squares fit.
}  
\label{scalelengths}
\end{figure}

The surprisingly good correlations between CO and K-band flux in disks
suggest that the scale lengths of the two distributions in the galaxy
disks might be similarly correlated.  
In Figure \ref{scalelengths}, we
plot the scale length of the stellar disk from the optical/near-infrared
image and the CO scale length determined from the low
resolution CO data.  
In both cases we fit the slope of the surface brightness ignoring the
central 50\arcsec\ to avoid the contribution from the bulge.
In Table \ref{scaletab} we list the scale lengths and 
uncertainties for each galaxy.
A weighted fit to the data yields a slope of
0.53$\pm$0.03 and an intercept of 0.90$\pm$.09,
with a correlation coefficient of 0.50, but there
is no systematic trend for either the CO or stellar disk to be longer
than the other because of the non-zero intercept of the fit.  This
non-zero intercept
is probably due to the large scatter in the data, which
exceeds the measurement uncertainties.  Another way to see this is to 
consider the mean ratio of the CO to stellar scale length, which is
0.88$\pm$0.14, with a dispersion of 0.52.  The uncertainty in the mean
is 16\%, even though the measurement uncertainty for any one galaxy is
typically less than about 10\%. 
Thus, it appears that the
stellar and CO scale lengths for the disks in the 15 galaxies plotted
are on average equal, but the relatively large scatter in the data
suggest that there are real variations in the ratio from galaxy to
galaxy.
The equality of the scale lengths, on
average, raises a question similar to that posed for the Milky Way
(e.g. Blitz 1996):
are the two scales lengths tied together, and if so, which controls
the other?  
Clearly, the scale lengths often differ 
in an individual galaxy (see especially NGC 628), contrary to 
the implication of the two-galaxy study by \cite{YS82}.

\section{Discussion}

\subsection{CO distributions at High and Low Resolution}

At 45\arcsec\ resolution, only 10 of the 193
galaxies surveyed as part of the
 FCRAO Extragalactic CO Survey \citep{Young95} exhibited rings, 
and 18 had peaks offset
from the nucleus. 
Thus 15\% of the FCRAO sample were inferred to have
distributions which lack a central peak.  
Our radial profiles derived from the single-dish data are consistent
with these results.
Of the 15 galaxies presented
in this paper, all show a single-dish radial profile which 
rises monotonically to the nucleus, supporting the idea that molecular gas
distributions which are not concentrated with the central
r $\la$ 5 kpc are relatively rare.

At full resolution, however, $\sim$50\% of
the SONG subsample galaxies do not exhibit peaks at the center.
This disparity with the low resolution results is unlikely to be due to the
selection of galaxies. 
The different conclusions
regarding central CO peaks derived from the FCRAO and SONG surveys are
unlikely to be due primarily to differences in the samples.  Indeed, if we
considered only our own (slightly lower than FCRAO resolution) 
12-meter data, we
would conclude that for nearly all galaxies the peak CO emission occurs at
the center. With the benefit of higher angular resolution, we find that
for at least 50\% of the SONG sample, the peak CO emission is not at the
center. In other words, with linear resolution on the order of a few 100
pc, only in roughly half the galaxies does CO peak at the center.
Thus, this analysis demonstrates the importance of
high-resolution observations of molecular gas and suggests that the 
conventional view that molecular gas generally peaks at the 
nucleus is incorrect.

\subsection{Radial Distribution of CO}

Probably the most striking feature in both the radial profile plots
(Figure \ref{coprofiles}) and the plots of CO-K offsets (Figure
\ref{alloffset} upper panel and middle panel) is the relationship between the
surface brightness of the bulge and the surface brightness of the
molecular gas as measured by the CO surface brightness.  
For all eight
galaxies that show an excess of CO
emission above the exponential disk profile in their central regions,
the excess is strikingly similar to the increase in the 
stellar light
due to the bulge component.  
The bulge component of the stellar radial
profile is generally attributed to a flattened spheroid.  On the other
hand, the molecular gas distribution almost certainly remains
disk-like; the gas is highly dissipative and the kinematics show 
regular rotation to first order. 
Thus, it is striking that the BIMA+12m profiles cannot, in general, be
fitted by a single exponential profile over all radii, and that
furthermore, many deviate from an exponential profile at approximately
the same radii as the central stellar bulge component starts to
dominate.  This is true not only for
strongly barred galaxies, such as NGC 2903 and NGC 3627, but also for
unbarred galaxies such as NGC 5055.

\subsection{ CO Emission Excesses in the Centers of Galaxies}

We consider two possibilities for the observed excess of CO
emission in the bulge region.
The two are not mutually exclusive and may, in fact, be related.
The first involves the hydrostatic pressure of gas in the bulge. 
The second, assume that the bulge has, in fact, formed from the molecular gas in
the nuclear regions of the galaxy.

For the bulge stars to be able to influence the
molecular gas near the galaxy center, there must be a mechanism to
transport information about the gravitational potential to the
molecular disk.  If there were a hot gas in the bulge region of these
galaxies with a filling factor approaching one, then it would provide
a mechanism for influencing the disk gas by increasing the pressure in
the bulge region.  \citet{SB92} show that the observed X-ray emitting
gas in the Milky Way will translate the gravitational potential of the
bulge into an increased pressure in the center of the Milky Way.  
This increased pressure leads to an increase in the average density of the
gas compared to the local value \citep{HB97b} and to an increase in
the internal velocity dispersion of the clouds \citep{Bally87}.  
This increased
density has also been seen in the bulge regions of other external
galaxies \citep{HB97a}.  Given the different environment in the bulge
regions, could this influence the conversion factor of CO luminosity
to H$_2$ mass, $X$, which has been determined locally?  

From the analysis of \cite{MB88} there are three
interacting effects that will change the 
relationship between
 CO luminosity
and H$_2$ mass: kinetic temperature, $T$, density, $\rho$, and velocity
dispersion increase over virial.  The interaction is such that
\begin{equation}
X=T^{-1}\rho^{1/2}(\sigma_{virial}/\sigma_{rms})^2. 
\end{equation}
A
decrease in $X$ indicates that the CO is more luminous for the same H$_2$ mass.
If we look at some typical values in the center of the Milky Way, the
clouds are five times hotter, 10 times denser, and have velocity
dispersions around 10 times higher than local clouds \citep{Bally87}.
This
increase in the density by a factor of 10 leads to an increase of a
factor of $\sqrt{10}$ in the virial velocity dispersion. Therefore, we
get X$_{bulge}$/X$_{local}$=0.06.

In other words, the conditions in regions like the Galactic center
would cause the CO to be overluminous by a factor
of 16, or 3 magnitudes, which is not far from the 2.2 magnitudes
that would be required for the molecular gas surface density to follow
the disk stellar surface brightness ($\S$\ref{cokratio}).  
This is in qualitative agreement with the 
deficit of gamma-rays at the center of the Milky Way \citep{Blitz85}
as well as the low value of
$X$ derived from dust measurements at the center of the Milky Way \citep{Sod95} and
in three BIMA SONG galaxies \citep{Regan00}.

It is worth noting that two (NGC 4414 and NGC 628) 
of the three galaxies with the latest Hubble
type show no
increase in surface brightness above an exponential profile in their
centers.
The
galaxy with the latest Hubble type in the subsample, NGC 6946, does
show an excess but it is also undergoing a nuclear starburst which
could provide an excess pressure through a large number of
supernovae. 
\cite{SB92} predicted that the smaller bulges of late type
galaxies would give rise to lower interstellar gas pressures than
earlier types.  
This would lead us
to expect that later Hubble type galaxies would not have as large of an
excess of CO emission in their central regions.

This interpretation of the cause of the excess CO emission in the
nuclear region of some of these galaxies could be tested by high
resolution millimeter observations of individual clouds in the bulge
regions of one of the galaxies with a central excess.  If the clouds
show a large internal velocity dispersion, similar to clouds in the
bulge region of the Milky Way, it could account for their brighter
central emission.

Another explanation for a concomitant increase in the CO and
stellar surface brightness in the nuclear regions is related to a
suggestion by \cite{Kormendy93} that at least some bulges are really
disk-like (``pseudo-bulges'') and form from the nuclear disk gas.  These
bulges have much smaller velocity dispersions than predicted by the
Faber-Jackson relation \citep{FabJack76} and relatively rapid rotation, suggesting
that they are rotationally flattened.  One of the prototype disk-like
bulges mentioned by \cite{Kormendy93} is NGC 4736, one of the galaxies in our
sample that shows a nuclear increase in the CO surface brightness in
the bulge region.  In this scenario, gas is brought to the center by
means of a bar or some other transport mechanism and, if the gas is
continually supplied, then the stars that form will have disk
kinematics.  
If these stars are efficiently scattered by either a bar,
molecular clouds in the disk, or a
bending instability \citep{Raha91,CombesSanders81} 
then they can form themselves
into the high angular momentum bulge that is seen in some cases.
Incidentally, the high pressures in the bulge region will continue to
keep the gas substantially molecular, as discussed above, which can
permit continued star formation as long as the gas supply holds out.
If a steady state is reached, then it would be natural for the K band
to CO surface densities to have values similar to the disks; all that
would be required is a near-constant star formation efficiency from the
molecular gas in both regions.

        One straightforward test of this hypothesis for the galaxies
in our sample would require that all of the galaxies that show a CO
excess in the bulge region ought to exhibit relatively high angular
momentum pseudo-bulges.  A second test would show metallicities
inconsistent with a closed-box evolutionary model of the star formation
for these bulges.

\subsection{Comparison with Previous Results}

A good agreement between the large scale radial distribution of the CO 
gas and the stellar disk was first noted by \cite{YS82} for NGC 6946
and IC 342.
Their under-sampled major and minor axis profiles (with an effective
resolution of $>$ 50\arcsec) were in very good
agreement with optical B-band radial profiles.
Although the central pointing of their observations
did show an excess similar to what 
our profiles show,
they concentrated their
discussion on the agreement between the scale length of the CO profiles 
(excluding the central pointing) and the B-band disk scale lengths.
Even at their 50\arcsec\ resolution, the IC 342 CO scale length agreed 
better with the B-band scale length 
when they excluded the inner three pointings, showing that the excess central 
emission affects even single dish profiles.
Our results in Figure \ref{scalelengths} show considerably more scatter
in the relationship between the scale lengths of the CO and
stellar disks than those suggested by \cite{YS82}.

Observations out to a radius of 30\arcsec\ of galaxies using the OVRO and NRO
interferometers were made by \citet{Sak99a}, who
found that the CO radial profiles decreased at a rate such that
the surface brightness is 1/$e$ of the central surface brightness 
at an average distance of 500 pc.
This is a much shorter distance than that found by single
dish observations \citep{YS82,YS91,Young95}.
In addition, by requiring a minimum surface brightness in
the FCRAO survey, the \citet{Sak99a} sample may have emphasized
galaxies with centrally concentrated CO distributions.
By comparing the \citet{Sak99a} results
with the radial profiles presented in Figures \ref{coprofiles}
and \ref{modprof} of this paper, it is apparent that in such
centrally concentrated galaxies, the region enclosed by
$r$=30\arcsec\ is essentially the bulge region that
we find to be characterized by an excess of emission over the
stellar disk (\S\ref{molgasdist}).  While it is possible to
represent the CO by an exponential in this nuclear region,
it is probably completely decoupled from the larger-scale exponential
that characterizes the stellar disk.

Previous single dish studies have also suggested that the CO surface
brightness varies by only a factor of two at a given radius
\citep{YS82,YS91}.  In contrast, the high-resolution BIMA SONG maps
demonstrate azimuthal variations of greater than 10 in many galaxies
(for example: NGC 5194, NGC 6946, NGC 628).  The distribution on small scales is
inherently different than the stars in that the CO is much more
clumpy.  
Although variations in the stellar surface brightness due to
non-axisymmetric features in the disk (e.g., bars and spiral arms) can
be quite strong in the vicinity of such features ($\sim$ a factor of 2), 
when the emission is averaged in annuli, the stellar
radial profiles show only slight variations from an exponential
profile.  
On the other hand, the CO distributions in most galaxies
show large variations from  exponential
profiles; 
as the molecular gas is dynamically cold
it responds strongly to perturbations such as bars,
spiral arms, and resonances.

\subsection{Spatial and Temporal Variations in the Disk Distribution}

The average ratio of the scale lengths of the stellar
and CO disks among the 15 galaxies in our sample is close to
unity, suggesting that the
two components are closely coupled.
This agreement may be a sign that the scale lengths of both components
averaged over time are the same within each individual galaxy.
Even though the
ratio of  CO and stellar light scale lengths shows substantial differences
between individual galaxies, a sample average close to unity suggests that
these differences average out over time, perhaps through some
feedback mechanism that suppresses the  variations.
One feedback mechanism that originates in the molecular gas might
be the formation of new stars.
As new stars form from the molecular gas they would have a radial
distribution similar to the molecular gas.
Over time, even if the the stellar disk started with a different
scale length, it would tend toward a scale length similar to that of
the molecular gas.
If this were true, one prediction is that the scale of a tracer
of current star formation such as H$\alpha$ or U-band would have
a scale length closer to the molecular gas scale length than the
stellar scale length.
Alternatively, if the gravitational potential of the stellar
disk exerted some feedback on the molecular disk, the stellar disk could
damp out variations in the molecular disk.  In this way, the stars in the
disk could prevent the molecular disk from diverging significantly from
the stellar radial profile.

The CO distributions are much more variable than the stellar disks,
in both the azimuthal and radial directions; 
this is a consequence of the lower random motions of
the molecular gas, which makes it respond more to gravitational
perturbations such as bars and spiral arms.
In addition, the molecular gas in a galaxy can be depleted by the
formation of new stars (either directly or by photodisassociation)
and can be created out of atomic gas.
Thus, while the stellar content of a galaxy can only increase with
time, the molecular gas content of a galaxy can rise and fall.
The combination of these two effects can explain the variable morphology
of these molecular distributions.

\section{Conclusions}

We introduced the BIMA Survey of Nearby Galaxies with a 
study of the CO radial brightness distributions of a subsample of 15
spiral galaxies.  Our conclusions are:

(1) The BIMA SONG maps display a variety of distributions of molecular
gas.  They are considerably more heterogeneous on sub-kiloparsec size
scales than are their stellar counterparts, both within a galaxy and
from source to source.

(2) The radial brightness distributions of CO emission reflect the
clumpy nature of the molecular gas distributions on size scales of a few hundred
parsecs.  This is unlike the stellar radial profiles, which tend to be
smooth even on small scales despite significant contrasts in the
azimuthal light distribution due to perturbations such as
bars and spiral arms.

(3) The CO radial profiles cannot be explained by a single-component
exponential disk model.  
In fact, in a majority of the galaxies the CO
profiles seem to follow the two-component (bulge + disk) stellar brightness
distribution quite well.  
The average ratio of the CO to stellar scale lengths in the
disks is 0.88$\pm$0.14, or close to unity, even though
the dispersion (0.52) in the ratio is large enough  to imply
real variations from galaxy to galaxy.  This may indicate that the
CO and stellar scale lengths for any given galaxy are the same
 when averaged over time.

(4) The ratios of the CO to K-band surface brightness 
in the galaxies as a whole, the bulge regions, and the disk regions show
a weak relationship with Hubble type.
The ratio of the bulges alone is only marginally larger than for the disks,
0.52$\pm$0.38 mag, much smaller than the value of 2.2 mag expected
from the inward extrapolation of the disk.
This implies a possible
physical connection between the the bulge and the inner CO surface brightness.

(5) The excess CO emission seen in the central regions of many of the
galaxies may either be due to an accumulation of molecular gas in the
nuclear region or it may
explained by an increase in the external pressure on the molecular
clouds in the bulge region from a hot diffuse gas.  This increased
pressure not only raises the density but raises the internal velocity
dispersion of the gas, increasing the CO luminosity relative to the
H$_2$ mass.

\acknowledgements 

The authors thank Rick Forster and the staff of the Hat Creek
Radio Observatory
at Hat Creek for their help with the BIMA observations, and we thank the
BIMA Board for dedicating so much time to this Observatory project.
We thank Peter Teuben for assistance with the software and for many
helpful discussions.  We thank Peter Guarnieri for his help with the
reduction of the optical data, and Annette Ferguson for providing the
R band images of NGC 628 and NGC 7331 for use in this analysis.
We would also like to thank the referee, Jeff Kenney, for his
helpful comments on the paper.
This research was supported by NSF
grants AST-9981308 (UC Berkeley), AST-9981289 (University of
Maryland), and AST-9900789 (Steward Observatory).


\singlespace
\clearpage
\epsscale{0.95}
\plotone{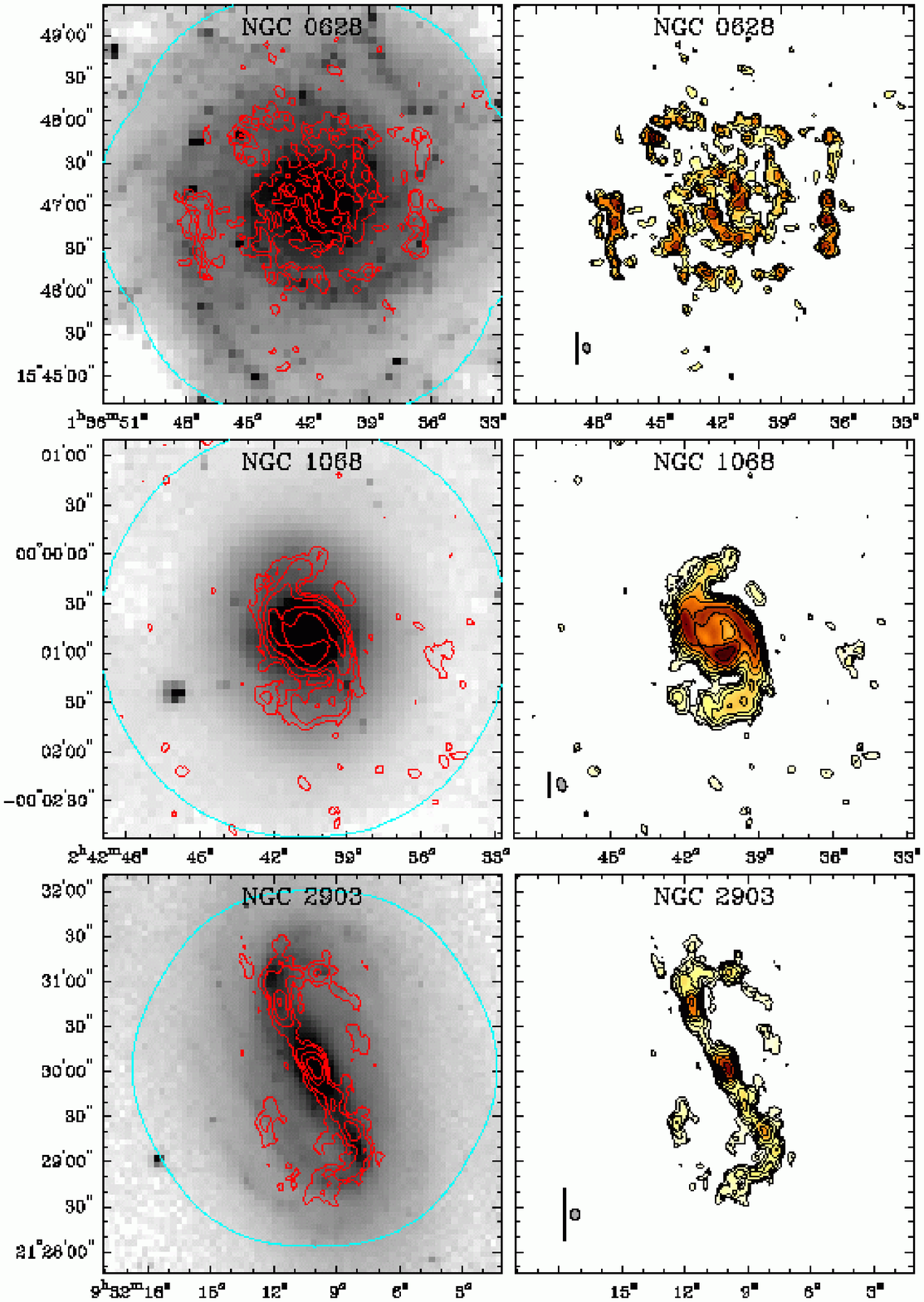}
\plotone{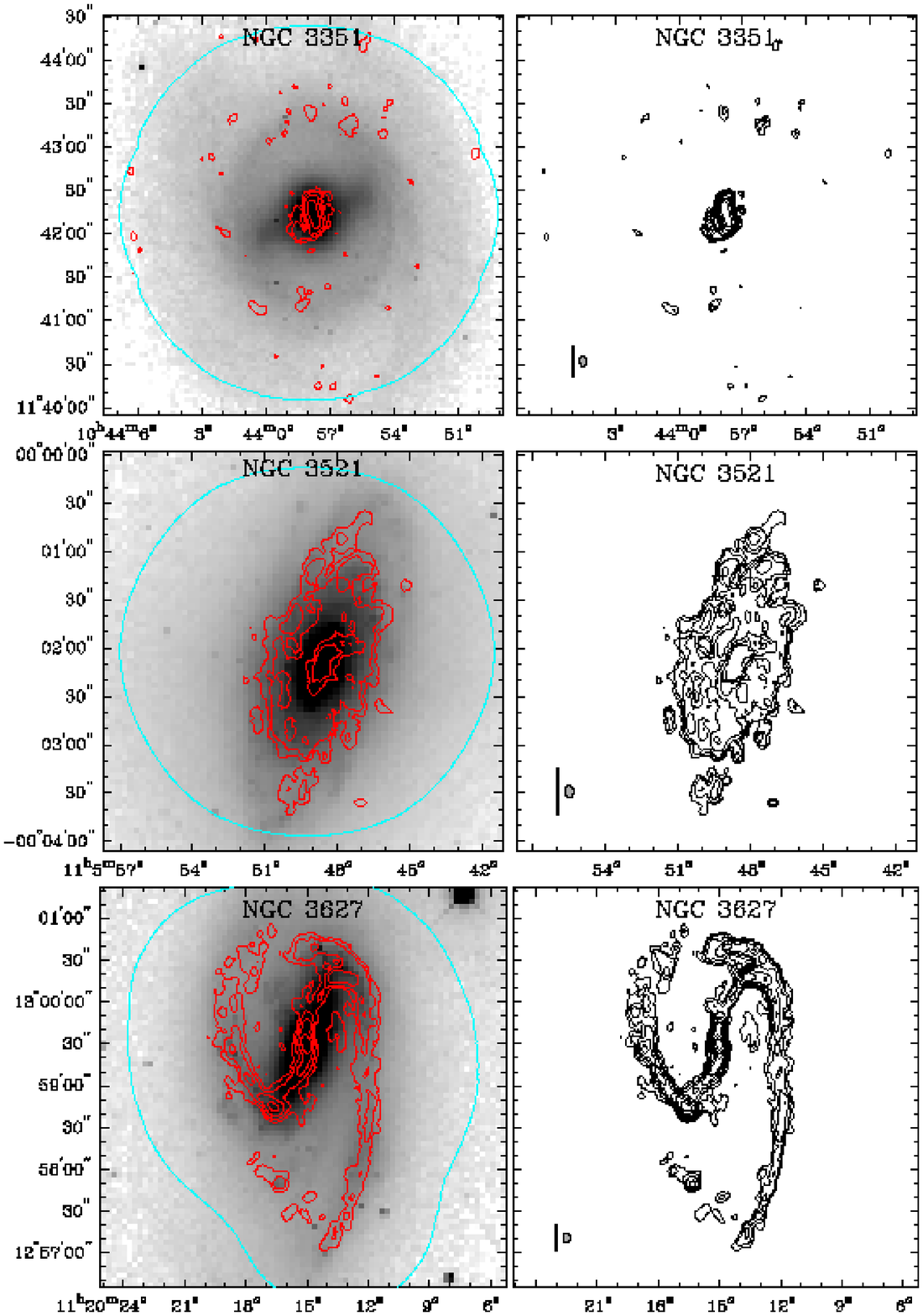}
\plotone{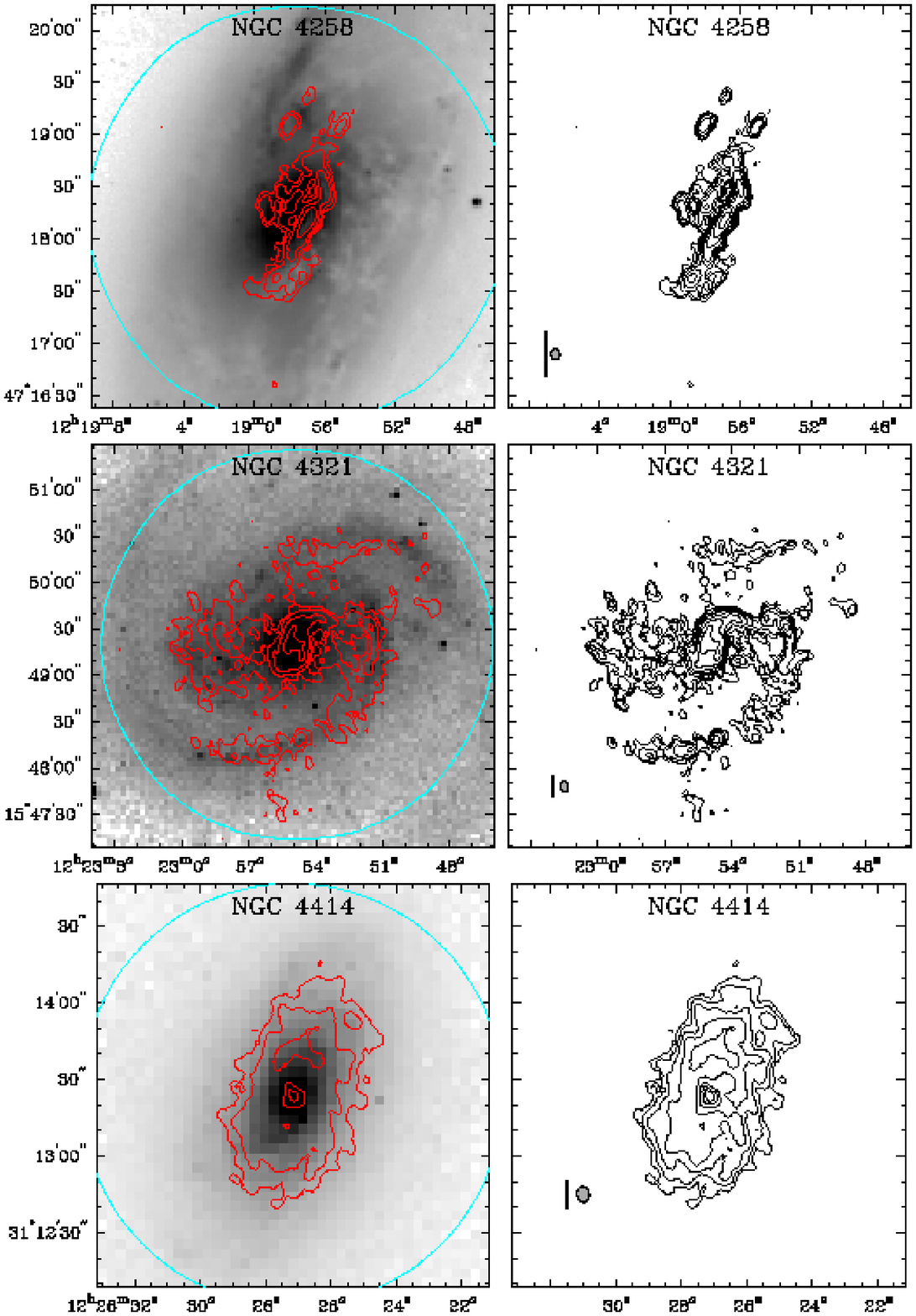}
\plotone{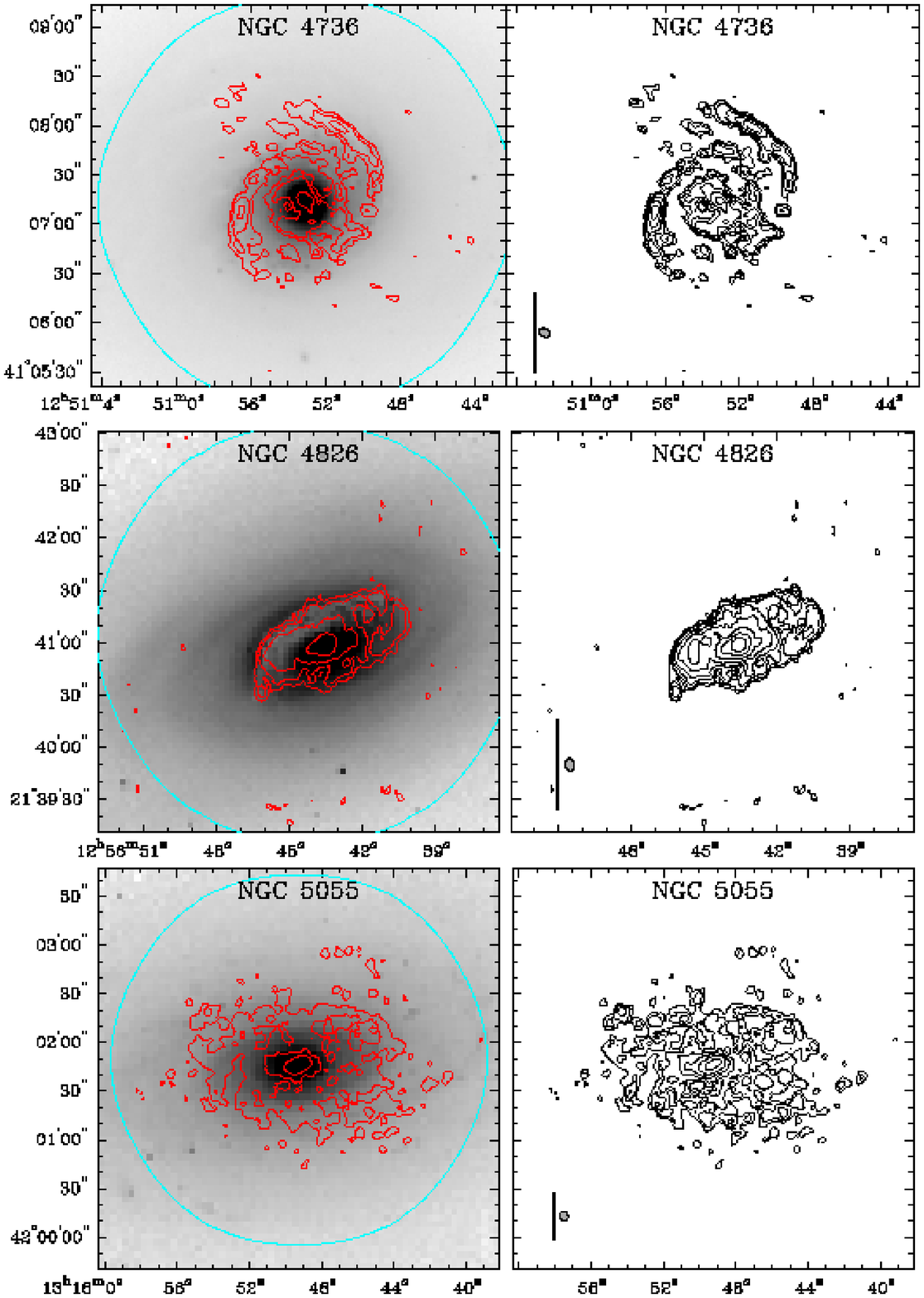}
\plotone{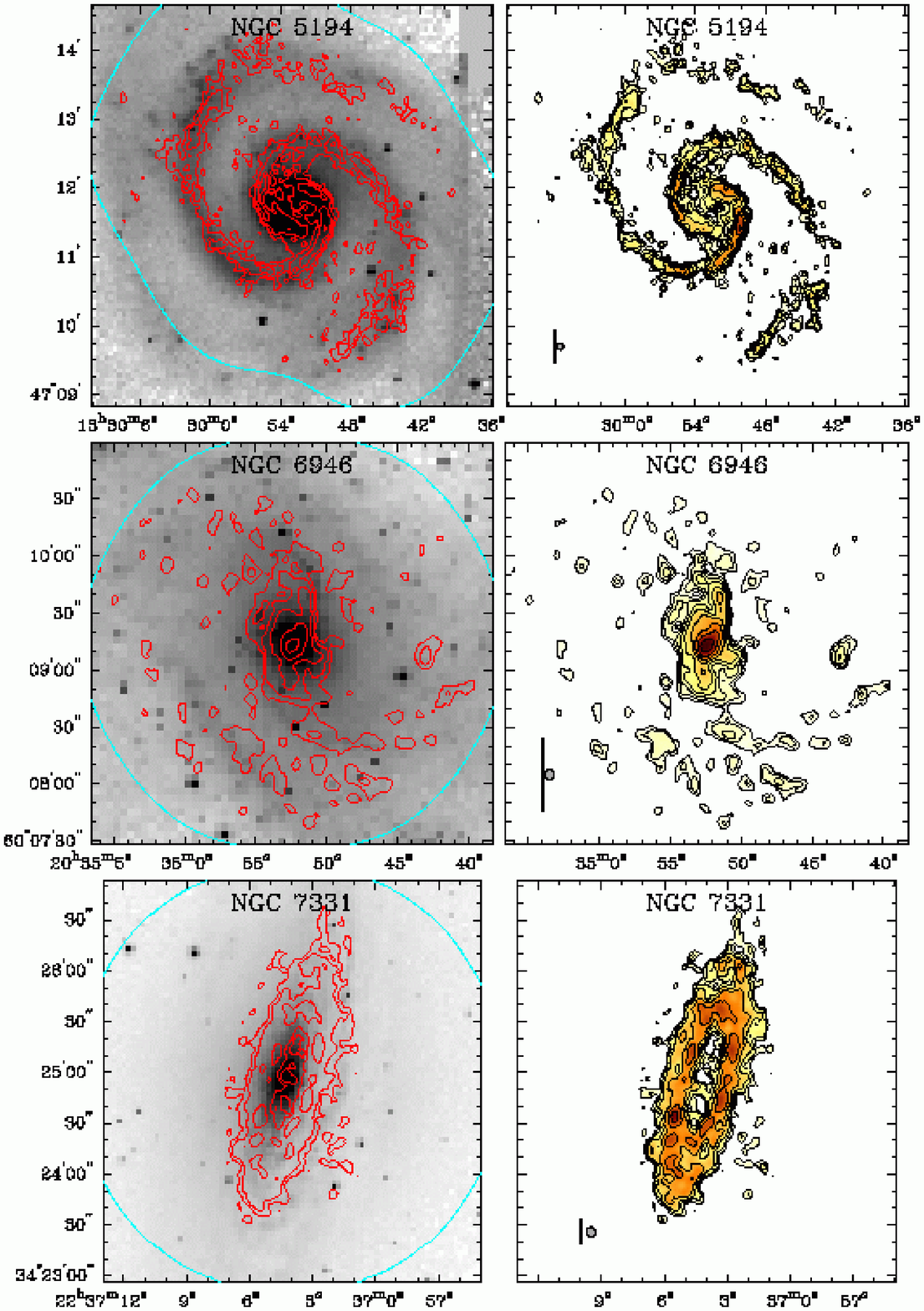}
\begin{figure}
\epsscale{0.5}
\caption{(left-hand column)
CO contours overlaid on stellar image.
The observing filter of the underlying images are
listed in Table~\ref{Toptir}. Each contour is at a surface brightness of
2.5119 times higher (one magnitude) than the previous one.  The
starting contour levels for the galaxies are: NGC 0628 - 1.5, NGC 1068
- 6, NGC 2903 - 5, NGC 3351 - 2.5, NGC 3521 - 5, NGC 3627 - 5, NGC 4258
- 4, NGC 4321 - 3, NGC 4414 - 3, NGC 4736 - 4, NGC 4826 - 4, NGC 5055
- 2, NGC 5194 - 5, NGC 6946 - 6, NGC 7331 - 3 Jy beam\sso\ km\
sec\sso.
(right-hand column)
Total intensity maps of sample galaxies. For each galaxy the vertical
bar in the lower left corner shows the angular size of one kpc at the
assumed distance to the galaxy.  The synthesized beam is shown next to
the kpc bar.  For all the galaxies the contours are spaced at
logarithmic intervals. Each contour is at a surface brightness of
1.585 times higher (one-half magnitude) than the previous one.  The
starting contour levels for the galaxies are the same as in the left-hand
column.}
\label{moments}
\end{figure}

\clearpage

\epsscale{0.65}
\plotone{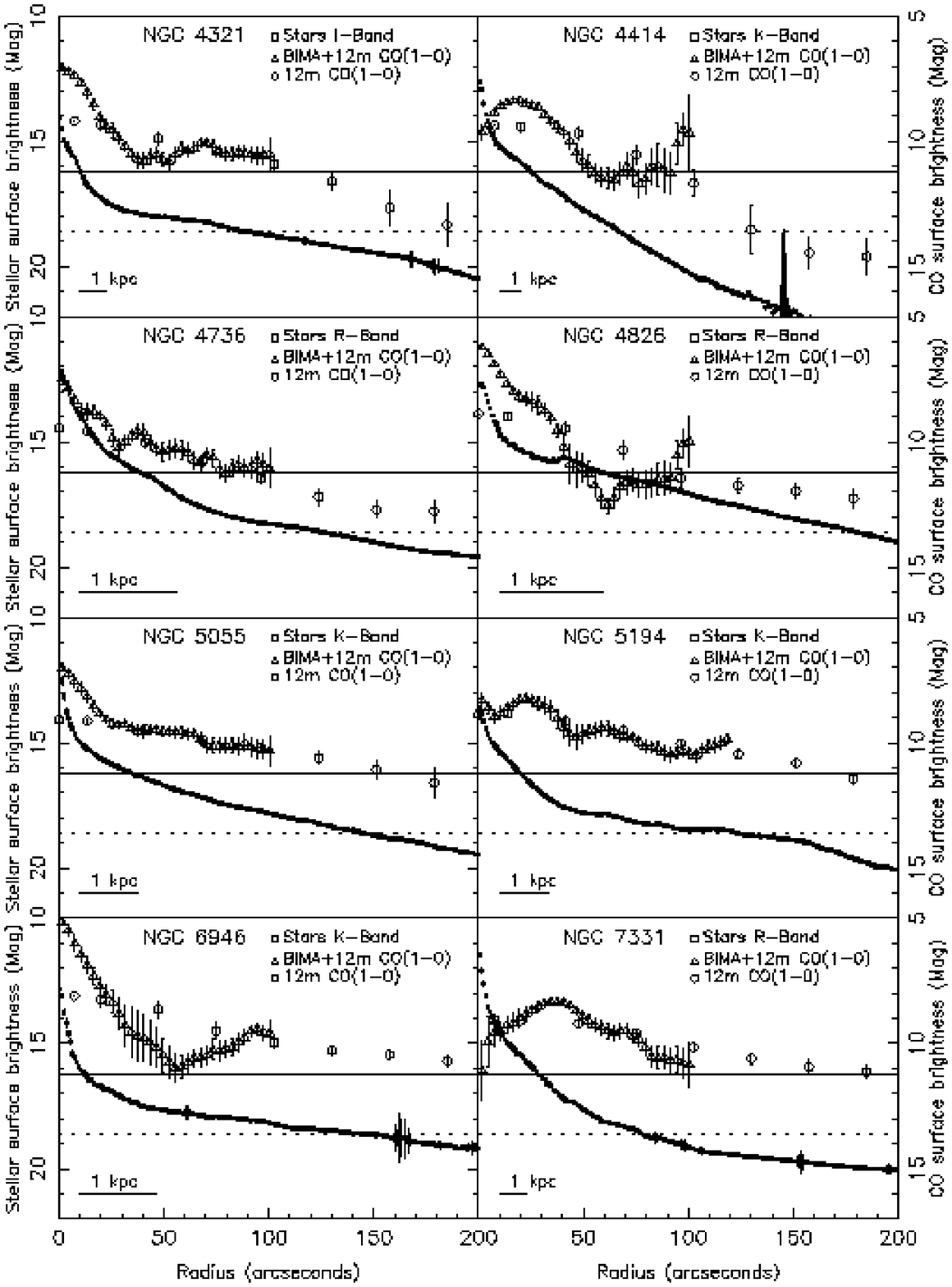}

\plotone{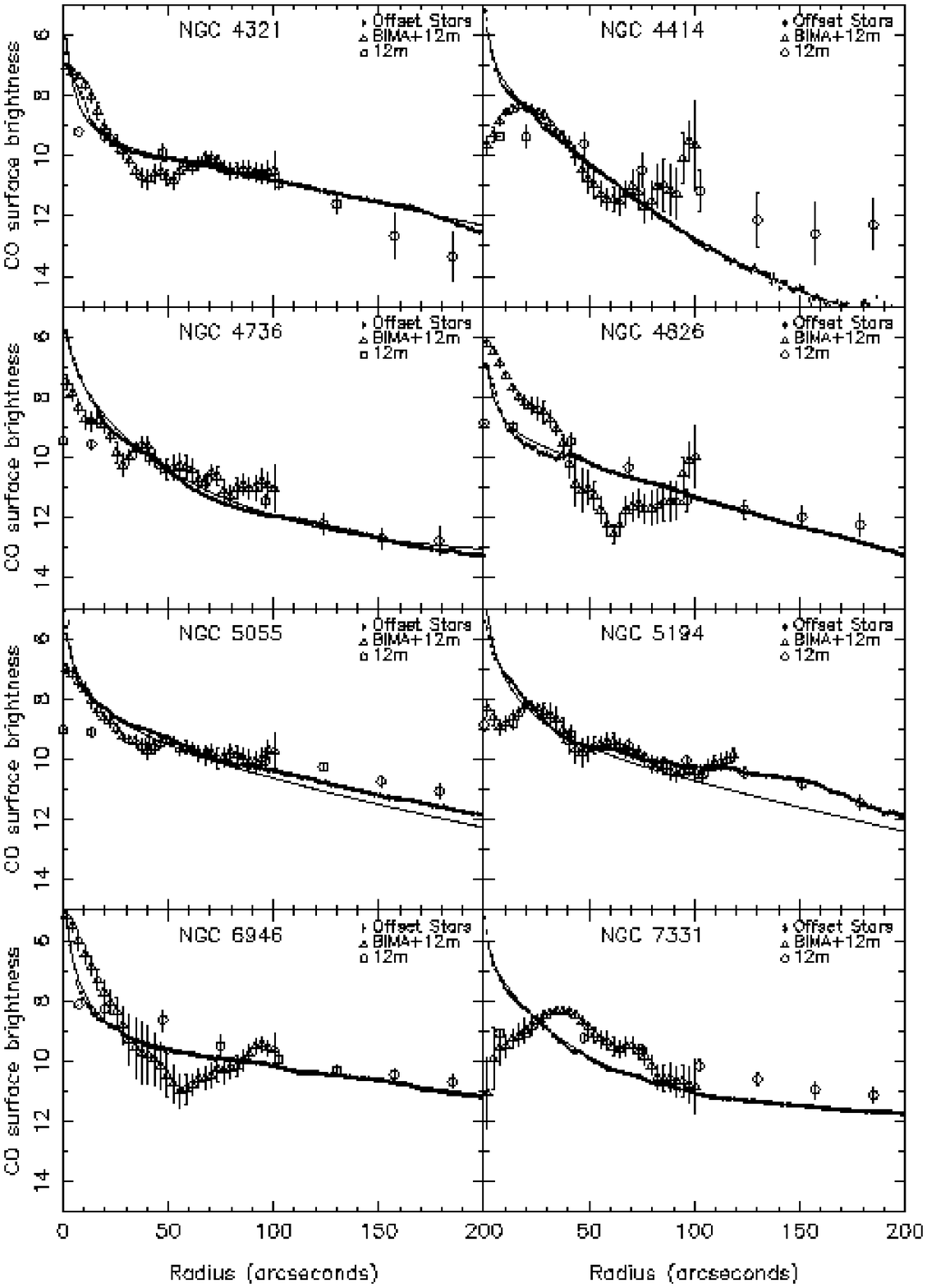}
\epsscale{0.5}

\clearpage


\begin{table}
\begin{center}
\caption{BIMA SONG Sample \label{sampletab}}
\begin{tabular}{lllllll}
\tableline\tableline
NGC 0628 & NGC 0925 & NGC 1068 & IC 342  & NGC 2403 & NGC 2841 & NGC 2903 \\
NGC 2976 & NGC 3031 & NGC 3184 & NGC 3344 & NGC 3351 & NGC 3368 & NGC 3521 \\
NGC 3627 & NGC 3726 & NGC 3938 & NGC 3953 & NGC 3992 & NGC 4051 & NGC 4258 \\
NGC 4303 & NGC 4321 & NGC 4414 & NGC 4450 & NGC 4490 & NGC 4535 & NGC 4548 \\
NGC 4559 & NGC 4569 & NGC 4579 & NGC 4699 & NGC 4725 & NGC 4736 & NGC 4826 \\
NGC 5005 & NGC 5033 & NGC 5055 & NGC 5194 & NGC 5248 & NGC 5247 & NGC 5457 \\
NGC 6946 & NGC 7331 & & & & & \\
\tableline
\end{tabular}
\end{center}
\end{table}

\clearpage

\begin{deluxetable}{lllrlrlrcc}
\tabletypesize{\scriptsize}
\tablecaption{Paper I Subsample\label{galparm}}
\tablehead{
\colhead{Galaxy} &
\colhead{RA (J2000)} &
\colhead{Dec (J2000)} &
\colhead{$V_{\rm LSR}$} &
\colhead{$i$ } &
\colhead{PA } &
\colhead{Type} &
\colhead{$d$} &
\colhead{Distance}&
\colhead{CO beam size}
\\
\omit &
\colhead{$h~~m~~s$} &
\colhead{\arcdeg~~\arcmin~~\arcsec} &
\colhead{km~s$^{-1}$} &
\colhead{\arcdeg} &
\colhead{\arcdeg} &
\omit &
\colhead{Mpc} &
\colhead{Reference}&
\colhead{\arcsec}
} 

\startdata
NGC 0628 & 01:36:41.70 & +15:46:59.4 & 657  & 24 & 25  & SA(s)c 
	 & 7.3  & 1 & 7.1$\times$5.2\\
NGC 1068 & 02:42:40.74 & $-$00:00:47.7 & 1136 & 33 & 13  & (R)SA(rs)b, Sy2  
	 & 14.4 & 1& 8.9$\times$5.6\\
NGC 2903 & 09:32:10.05 & +21:30:02.0 & 556  & 61 & 17  & SAB(rs)bc, HII   
	 & 6.3  & 1 & 6.8$\times$5.5\\
NGC 3351 & 10:43:57.98 & +11:42:14.4 & 778  & 40 & 13  & SB(r)b, HII      
	 & 10.1 & 2 & 7.3$\times$5.1\\
NGC 3521 & 11:05:49.26 & $-$00:02:02.3 & 805  & 58 & 164 & SAB(rs)bc, LINER 
	 & 7.2  & 1 & 8.7$\times$5.6\\
NGC 3627 & 11:20:15.07 & +12:59:21.7 & 727  & 63 & 176 & SAB(s)b, Sy      
	 & 11.1 & 3 & 6.6$\times$5.5\\
NGC 4258 & 12:18:57.52 & +47:18:14.2 & 448  & 65 & 176 & SAB(s)bc, Sy1    
	 & 8.1  & 4 & 6.0$\times$5.3\\
NGC 4321 & 12:22:54.84 & +15:49:20.0 & 1571 & 30 & 154 & SAB(s)bc, HII    
	 & 16.1 & 5 & 7.2$\times$4.9\\ 
NGC 4414 & 12:26:27.19 & +31:13:24.0 & 716  & 55 & 159 & SA(rs)c?         
	 & 19.1 & 6 & 6.3$\times$5.0\\
NGC 4736 & 12:50:53.06 & +41:07:13.6 & 308  & 35 & 100 & (R)SA(r)ab,LINER 
	 & 4.3  & 1 & 6.9$\times$5.0 \\
NGC 4826 & 12:56:44.24 & +21:41:05.1 & 408  & 54 & 111 & (R)SA(rs)ab, Sy  
	 & 4.1  & 1 & 7.4$\times$5.1\\
NGC 5055 & 13:15:49.25 & +42:01:49.3 & 504  & 56 & 81 & SA(rs)bc,HII/LINER 
	 & 7.2  & 1 & 5.8$\times$5.4\\
NGC 5194 & 13:29:52.35 & +47:11:53.8 & 463  & 15 & 0  & SA(s)bc-pec, Sy2.5 
	 & 8.4  & 1 & 5.8$\times$5.1\\
NGC 6946 & 20:34:52.33 & +60:09:14.2 & 48   & 54 & 65  & SAB(rs)cd, HII  
	 & 5.5  & 1 & 5.9$\times$4.9\\
NGC 7331 & 22:37:04.09 & +34:24:56.3 & 821  & 62 & 172 & SA(s)b, LINER   
	 & 15.1 & 7 & 6.1$\times$5.0\\
\enddata
\tablerefs{
(1) \citet{Tully88};
(2) \citet{Graham97};
(3) \citet{Saha99};
(4) \citet{Maoz99};
(5) \citet{Ferrarese96};
(6) \citet{Turner98};
(7) \citet{Hughes98};
}
\end{deluxetable}

\clearpage	
		
\begin{deluxetable}{lllll}
\tabletypesize{\scriptsize}
\tablecaption{Optical/Infrared Data \label{Toptir}}
\tablehead{	
\colhead{Galaxy}  & 
\colhead{Band } & 
\colhead{image reference} & 
\colhead{X-K color} & 
\colhead{photometry reference} 
}
\startdata
NGC 0628 & R  & \citet{Ferguson98}       & 2.4  &\citet{Ar77} \\
NGC 1068 & J  & \citet{Regan00}          & 1.16 &\citet{Ar77}\\
NGC 2903 & K' & \citet{RE97}             &  --    &\\
NGC 3351 & I  & SONG complementary       & 1.65 &\citet{Tifft61,Glass76,Ar77}\\
NGC 3521 & K' & \citet{Thornley96}       & --     & \\
NGC 3627 & K' & \citet{RE97}             & --   & \\
NGC 4258 & R  & SONG complementary       & 2.35 & \citet{Ar77}\\
NGC 4321 & I  & SONG complementary       & 1.97 & \citet{BSS83,Ar77}\\
NGC 4414 & K' & \citet{Thornley96}       & --   & \\
NGC 4736 & R  & SONG complementary       & 2.17 &\citet{Johnson66}  \\
NGC 4826 & R  & SONG complementary       & --   &  \citet{Ar77}\\
NGC 5055 & K' & \citet{Thornley96}       & --   & \\
NGC 5194 & K  & \citet{Gruendl96}        & --   & \citet{Ar77,Ellis82}\\
NGC 6946 & K  & \citet{RV94}             & --   & \\
NGC 7331 & R  & A. Ferguson, priv. comm. & --   &\citet{Ar77}\\
\enddata
\end{deluxetable}

\clearpage
\begin{deluxetable}{lccc|ccc|ccc}
\tablecaption{Stellar and CO Magnitudes \label{KCOmag}}
\tablehead{
\colhead{Galaxy} & 
 \colhead{} &\colhead{Total} & \colhead{} &
 \colhead{} &\colhead{Bulge} & \colhead{} &
 \colhead{} &\colhead{Disk} & \colhead{} \\
\colhead{} &
\colhead{CO} & \colhead{K} & \colhead{CO-K} &
\colhead{CO} & \colhead{K} & \colhead{CO-K} &
\colhead{CO} & \colhead{K} &\colhead{CO-K} }
\startdata
NGC 0628 & $-$1.00&  6.78& $-$7.78 &  2.52&  9.31& $-$6.79 & $-$0.96&  6.89& $-$7.85\\
NGC 1068 & $-$1.45&  5.69& $-$7.14 &  0.86&  6.96& $-$6.10 & $-$1.32&  6.09& $-$7.40\\
NGC 2903 & $-$1.07&  6.01& $-$7.08 &  0.37&  7.71& $-$7.34 & $-$0.73&  6.26& $-$7.00\\
NGC 3351 & $-$0.13&  6.53& $-$6.66 &  2.09&  7.91& $-$5.81 &  0.02&  6.89& $-$6.87\\
NGC 3521 & $-$1.41&  5.82& $-$7.24 &  1.16&  7.21& $-$6.05 & $-$1.30&  6.18& $-$7.48\\
NGC 3627 & $-$1.31&  6.03& $-$7.34 &  0.73&  8.06& $-$7.32 & $-$1.13&  6.22& $-$7.35\\
NGC 4258 & $-$0.75&  5.89& $-$6.63 &  0.47&  8.04& $-$7.57 & $-$0.32&  6.05& $-$6.37\\
NGC 4321 & $-$1.16&  6.32& $-$7.48 &  0.97&  9.16& $-$8.19 & $-$1.00&  6.40& $-$7.40\\
NGC 4414 & $-$0.06&  6.83& $-$6.89 &  3.67&  8.74& $-$5.07 & $-$0.03&  7.03& $-$7.06\\
NGC 4736 & $-$0.86&  5.31& $-$6.17 &  0.16&  5.87& $-$5.71 & $-$0.32&  6.30& $-$6.62\\
NGC 4826 & $-$1.01&  5.04& $-$6.05 & $-$0.45&  6.26& $-$6.71 & $-$0.03&  5.46& $-$5.49\\
NGC 5055 & $-$1.52&  5.69& $-$7.21 &  0.65&  7.36& $-$6.71 & $-$1.36&  5.95& $-$7.31\\
NGC 5194 & $-$2.37&  5.72& $-$8.09 &  0.34&  7.55& $-$7.21 & $-$2.27&  5.95& $-$8.22\\
NGC 6946 & $-$2.13&  6.05& $-$8.17 & $-$0.88&  7.99& $-$8.87 & $-$1.71&  6.25& $-$7.96\\
NGC 7331 & $-$1.51&  5.86& $-$7.37 &  3.93&  7.94& $-$4.02 & $-$1.51&  6.03& $-$7.54\\
\enddata
\end{deluxetable}

\clearpage
\begin{deluxetable}{lccc}
\tablecaption{Stellar and CO profile parameters \label{scaletab}}
\tablehead{
\colhead{Galaxy} & \colhead{Stellar scale length} &
\colhead{CO scale length} & Central excess?\\
\colhead{} & \colhead{(kpc)} & \colhead{(kpc)} & }
\startdata
NGC 0628& 3.4$\pm$0.01& 5.8$\pm$0.2& no\\
NGC 1068& 3.7$\pm$0.1& 1.7$\pm$0.2 & no\\
NGC 2903& 1.7$\pm$0.01& 1.6$\pm$0.1 & yes\\
NGC 3351& 2.4$\pm$0.02& 2.6$\pm$0.3 & yes\\
NGC 3521& 1.7$\pm$0.01& 2.3$\pm$0.1 & no\\
NGC 3627& 3.5$\pm$0.02& 1.8$\pm$0.6 & yes\\
NGC 4258& 3.6$\pm$0.03& 2.5$\pm$0.08& no\\
NGC 4321& 5.6$\pm$0.04& 2.8$\pm$0.3 & yes\\
NGC 4414& 2.9$\pm$0.05& 4.6$\pm$0.91 & no\\
NGC 4736& 0.7$\pm$0.01& 1.0$\pm$0.1& yes\\
NGC 4826& 1.1$\pm$0.0030& 1.1$\pm$0.2& yes\\
NGC 5055& 2.2$\pm$0.01& 2.6$\pm$0.1 & yes\\
NGC 5194& 3.2$\pm$0.07& 2.8$\pm$0.1& no\\
NGC 6946& 2.9$\pm$0.02& 2.1$\pm$0.3 & yes\\
NGC 7331& 4.3$\pm$0.04& 5.6$\pm$0.4 & no\\
\enddata
\end{deluxetable}
\end{document}